\documentclass{webofc}
\usepackage[varg]{txfonts}   

\usepackage{color}
\usepackage[dvipsnames]{xcolor}
\usepackage{caption}
\usepackage{subcaption}
\usepackage{pifont}
\usepackage{hyperref}

\begin{document}

\title{Equivariance and generalization in neural networks}

\author{\firstname{Srinath} \lastname{Bulusu}\inst{1}\fnsep\thanks{\email{sbulusu@hep.itp.tuwien.ac.at}} \and \firstname{Matteo} \lastname{Favoni}\inst{1,2}\fnsep\thanks{\email{favoni@hep.itp.tuwien.ac.at}} \and \firstname{Andreas} \lastname{Ipp}\inst{1}\fnsep\thanks{\email{ipp@hep.itp.tuwien.ac.at}} \and \firstname{David I.} \lastname{M\"uller}\inst{1}\fnsep\thanks{\email{dmueller@hep.itp.tuwien.ac.at}} \and\firstname{Daniel} \lastname{Schuh}\inst{1}\fnsep\thanks{\email{schuh@hep.itp.tuwien.ac.at}}}

\institute{Institute for Theoretical Physics, TU Wien, \\
	Wiedner Hauptstr. 8-10, 1040 Vienna, Austria
\and Speaker and corresponding author}

\abstract{The crucial role played by the underlying symmetries of high energy physics and lattice field theories calls for the implementation of such symmetries in the neural network architectures that are applied to the physical system under consideration. In these proceedings, we focus on the consequences of incorporating translational equivariance among the network properties, particularly in terms of performance and generalization. The benefits of equivariant networks are exemplified by studying a complex scalar field theory, on which various regression and classification tasks are examined. For a meaningful comparison, promising equivariant and non-equivariant architectures are identified by means of a systematic search. The results indicate that in most of the tasks our best equivariant architectures can perform and generalize significantly better than their non-equivariant counterparts, which applies not only to physical parameters beyond those represented in the training set, but also to different lattice sizes.}

\maketitle

\section{Introduction}

The past century has firmly established symmetries as a cornerstone of physics and especially of field theories, tracing back to Noether's theorem~\cite{Noether:1918}. Later on, the importance of local symmetries was understood, which then culminated in the development of the Standard Model of particle physics~\cite{Weinberg:1968,Salam:1968}, where the fundamental interactions are described as gauge theories.

In the past decade, machine learning techniques have been sharpened a great deal in many areas, especially in computer vision, and their success has attracted a lot of interest from the physics community, which progressively assimilated them into its arsenal of methods. In particular, neural networks (NNs) proved to be a very effective tool in phase detection~\cite{Carrasquilla:2017} and regression tasks on observables and physical parameters~\cite{Boram:2019,Shanahan:2018}, but in many cases no structural adaptation from image processing to physical problems was put in place. Bearing in mind the central role of symmetries, one should take them into account ab initio when building a fitting architecture for the theory in question. This means that, from a theoretical standpoint, a natural approach is to design the network architectures so that they inherently respect the desired symmetries. Whether a symmetry-preserving construction is beneficial in terms of actual performance can be problem dependent, but there have been recent indications that this is a promising strategy. In~\cite{Cohen:2016}, group equivariant convolutional neural networks (G-CNNs) are proposed, which take care of the incorporation of translational, rotational and reflection symmetry in the architecture. The extension of this approach to local symmetries is formalized in~\cite{Cohen:2019}. Other proposals have been made in the context of lattice field theory, for example by designing a gauge equivariant normalizing flow~\cite{Kanwar:2020,Boyda:2020} and with the creation of specific layers that preserve gauge equivariance~\cite{Favoni:2020}.
In these proceedings, we focus on invariance under spacetime translations and study the impact of different choices of architectures that do or do not break translational symmetry, investigating in particular the generalization to different physical parameters and lattice sizes, as was discussed in our paper~\cite{Bulusu:2021rqz}.

\section{Architecture choices}

As mentioned, G-CNNs deal with all the spacetime isometries, while here our interest is restricted to translations only. A more natural choice is represented by convolutional neural networks (CNNs), which were precisely conceived to impose translational symmetry in NNs.

\begin{figure}[t]
\centering
\begin{subfigure}[b]{0.45\textwidth}
\includegraphics[scale=0.7]{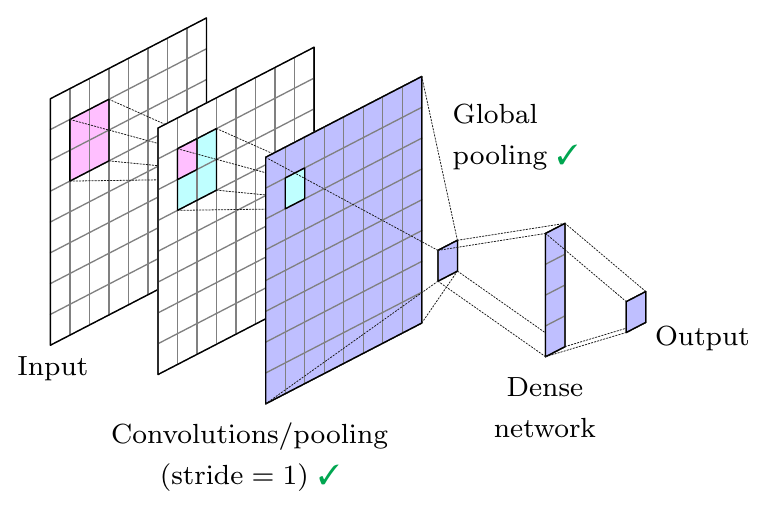}
\caption{Equivariant architecture (EQ)}
\end{subfigure}
\hfill
\begin{subfigure}[b]{0.45\textwidth}
\includegraphics[scale=0.7]{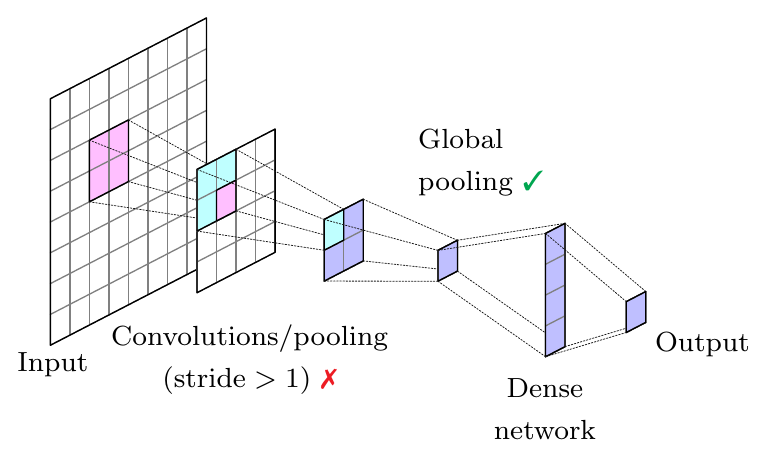}
\caption{Strided architecture (ST)}
\end{subfigure}

\centering
\begin{subfigure}[h]{0.45\textwidth}
\includegraphics[scale=0.7]{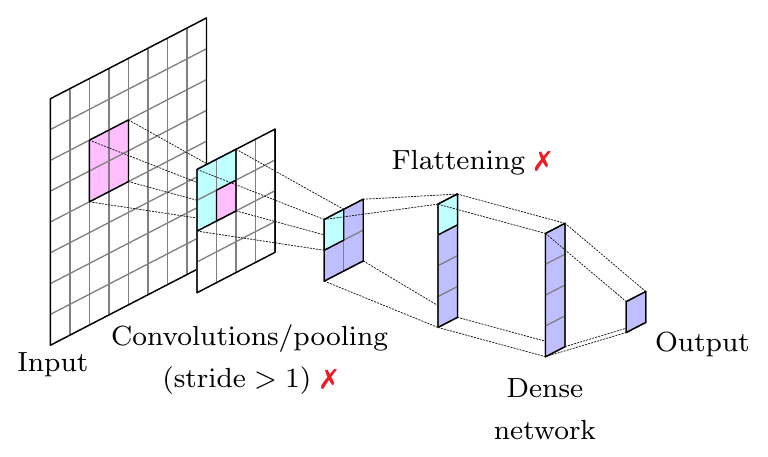}
\caption{Flattening architecture (FL)}
\end{subfigure}
\caption{The architecture types employed in this study. Translational equivariance is preserved by layers denoted by \textcolor{Green}{\ding{51}} and broken by layers marked with \textcolor{Red}{\ding{55}}. Convolutions and spatial pooling layers with a stride of one respect equivariance (a), whereas a stride larger than one or a flattening layer ruin it, as in (b) and (c). Flattening (c) also forces the architecture to be used on only one specific lattice size. Figures from~\cite{Bulusu:2021rqz}.}
\label{fig:archs}
\end{figure}

In order to study the influence of translational equivariance, we set up three different kinds of architectures, as depicted in fig.~\ref{fig:archs}. A stride of one in the convolutional and spatial pooling layers guarantees that translational symmetry is not broken in the first architecture (EQ). The symmetry is exact if one uses circular padding, which is the case if the theory features periodic boundary conditions. It is also not broken by the particular type of global pooling chosen, nor by the optional linear layers that are appended afterwards. The second architecture (ST) breaks equivariance because of a stride larger than one in the convolutional part. A residual symmetry based on translations that are a multiple of the stride holds, but general equivariance is not satisfied. In the third architecture (FL), the global pooling layer is replaced by a flattening step, which breaks the symmetry completely and restricts the use of the architecture to a specific lattice size.

These three types of networks are typical examples of architectures used in literature.

\section{Physical model}

Given the previous discussion, a good candidate to test our considerations is a relatively simple model characterized by translational symmetry and periodic boundary conditions. We opt for a complex scalar field in 1+1 dimensions with nonzero chemical potential $\mu$ and quartic interaction, also because of a pre-existing study to make comparisons with~\cite{Zhou:2019}. The Euclidean lattice action has the following expression:
\begin{equation}
    S_{lat}=\sum_x \left( \eta \lvert \phi_x \rvert^2 + \lambda \lvert \phi_x \rvert^4 -\sum_{\nu = 1}^2 \left( e^{\mu \mspace{2mu} \delta_{\nu, 2}} \phi_x^* \phi_{x + \hat{\nu}} + e^{- \mu \mspace{2mu} \delta_{\nu, 2}} \phi_x^* \phi_{x - \hat{\nu}} \right) \right),
\end{equation}
with $\eta=4+m^2$. A sign problem arises as a consequence of the complex terms originating from the products of fields at different lattice sites. It can be removed, though, thanks to a dual formulation, called flux representation \cite{Gattringer:2013b}, which maps the field $\phi_x$ into the integer fields $k_{x,\nu}$ and $l_{x,\nu}$, with $\nu$ indicating the temporal or the spatial direction. Configurations of these fields are generated as in~\cite{Gattringer:2013b} and are the input of the NNs.

\section{Prediction of observables}

We tackle now a regression task, where the networks have to learn the relationship between a lattice configuration and two observables, which was already carried out in~\cite{Zhou:2019}. The observables are the particle number density and the average of the modulo squared field, respectively given by
\begin{equation}
	n=\frac{1}{N}\sum_xk_{x,2}\,,\quad |\phi|^2=\frac{1}{N}\sum_x\frac{W(f_x+2)}{W(f_x)},
\end{equation}
with $
    f_x=\sum_\nu[|k_{x,\nu}|+|k_{x-\hat{\nu},\nu}|+2(l_{x,\nu}+l_{x-\hat{\nu},\nu})]\,$ and $W(f_x)=\int_0^\infty \mathrm{d}x\, x^{f_x+1}\mathrm{e}^{-\eta x^2-\lambda x^4}$.
Given the intensive nature of these observables, a global average pooling at the end of the convolutional part is the most appropriate choice.

\begin{figure}[!t]
    \minipage{0.47\textwidth}
    \includegraphics[scale=.75]{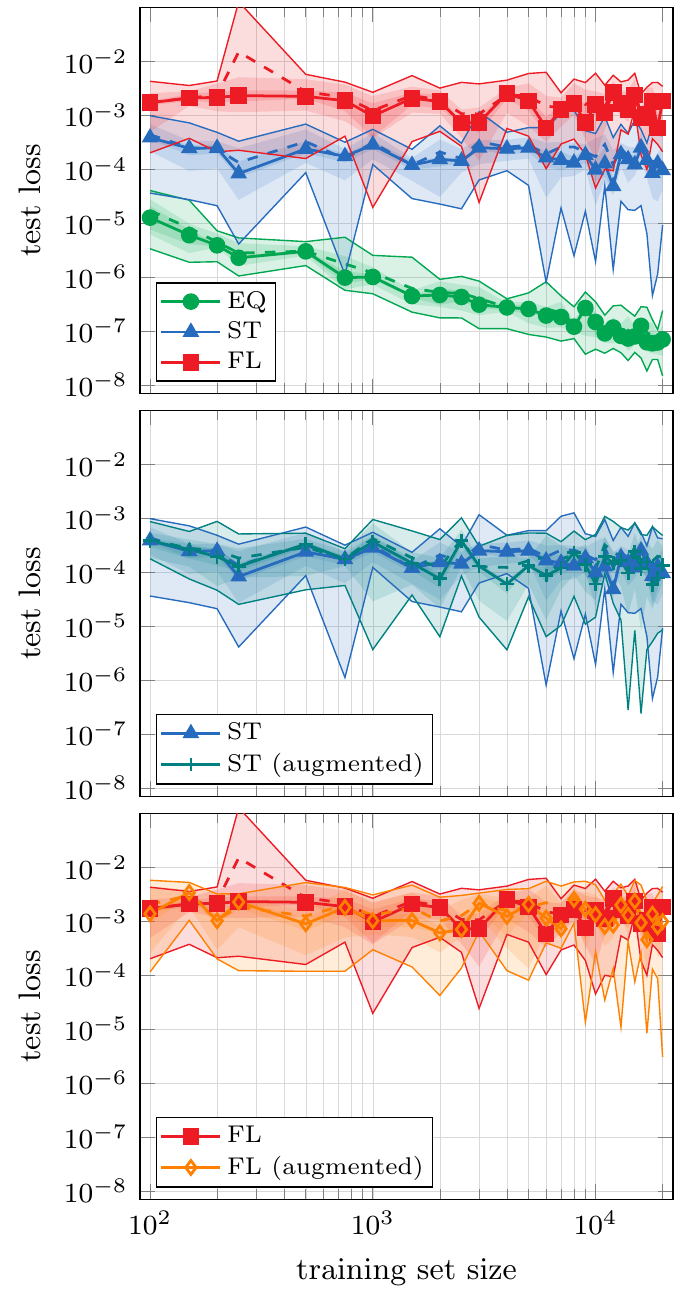}
    \caption{Performance comparison of best models suggested by Optuna. Ten instances with the most promising hyperparameter combination are trained at several training set sizes for each architecture type. The bands are delimited by the highest and lowest test loss, the dashed lines denote the average test losses and the continuous lines pass through the median test losses. The top plot shows the comparison of the three architecture types with no data augmentation, the middle and the bottom ones feature the test loss values with data augmentation for ST and FL, respectively. Image from~\cite{Bulusu:2021rqz}.}
    \label{fig:loss_vs_samples}
    \endminipage
    \hfill
    \minipage{0.47\textwidth}
    \includegraphics[scale=.7]{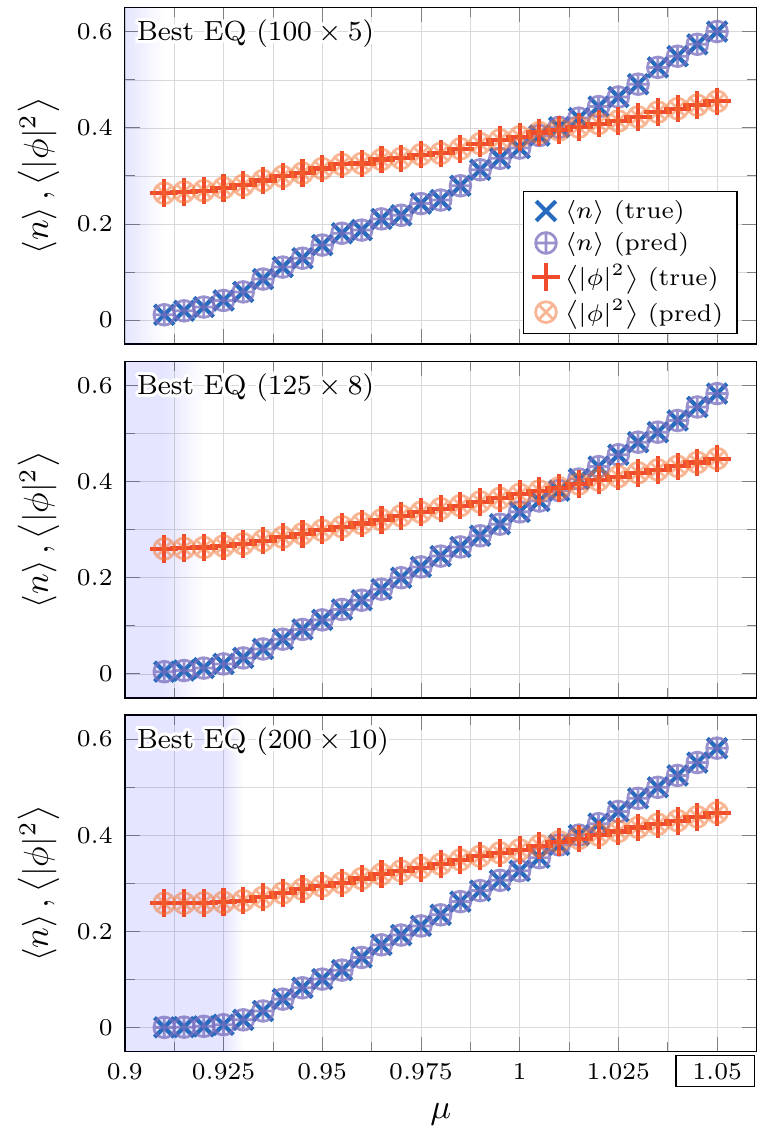}
    \caption{Silver blaze phase transition. The best EQ architecture is tested on chemical potentials and lattice sizes different from the ones it has been trained on (1.05 and $60\times4$, respectively). The silver blaze phase transition, which is better visible on larger lattices, is correctly predicted, even though training happened at only one phase. Image from~\cite{Bulusu:2021rqz}.}
    \label{fig:silver_blaze}
    \endminipage
\end{figure}

In order to drastically reduce human biases in the construction of specific architectures to compare, we employ an automatic hyperparameter optimizer, called Optuna~\cite{Akiba:2019}. We design a large search space in the three cases, allowing for networks ranging from very small to pretty large in terms of number of trainable parameters. We require Optuna to try the same hyperparameters three times in order to reduce the influence of the initialization. At the end of the search, we select the combination with the smallest average validation loss (which in this case is the mean squared error). Ten instances with such a combination are trained, and the procedure is repeated for each architecture and for various training set sizes. Tests are run on configurations with the same size used for training ($60\times4$), while the chemical potential ranges from 0.91 to 1.05 with steps of 0.005. Training takes place only at the uppermost value mentioned. The results are shown in fig.~\ref{fig:loss_vs_samples}. The top plot reports the behavior of the test loss as a function of the training set size for all three architectures. The EQ architecture clearly outperforms the non-equivariant types, and it improves its accuracy when more samples are provided during training, while the other two do not benefit significantly with more samples. Remarkably, this is also the case when using data augmentation while training, as shown in the middle and bottom plots.

Figure~\ref{fig:silver_blaze} shows the generalization capabilities of the EQ models. As said, training occurs on lattices with dimension $60\times4$ with chemical potential $\mu=1.05$, but the predictions remain very close to the true values also for different values of $\mu$ and of the lattice size. This even allows to detect the silver blaze phase transition~\cite{Gattringer:2013b} on the largest lattice even though training happened only in the disordered phase.

\begin{figure}[t]
    \centering
    \includegraphics[scale=0.8]{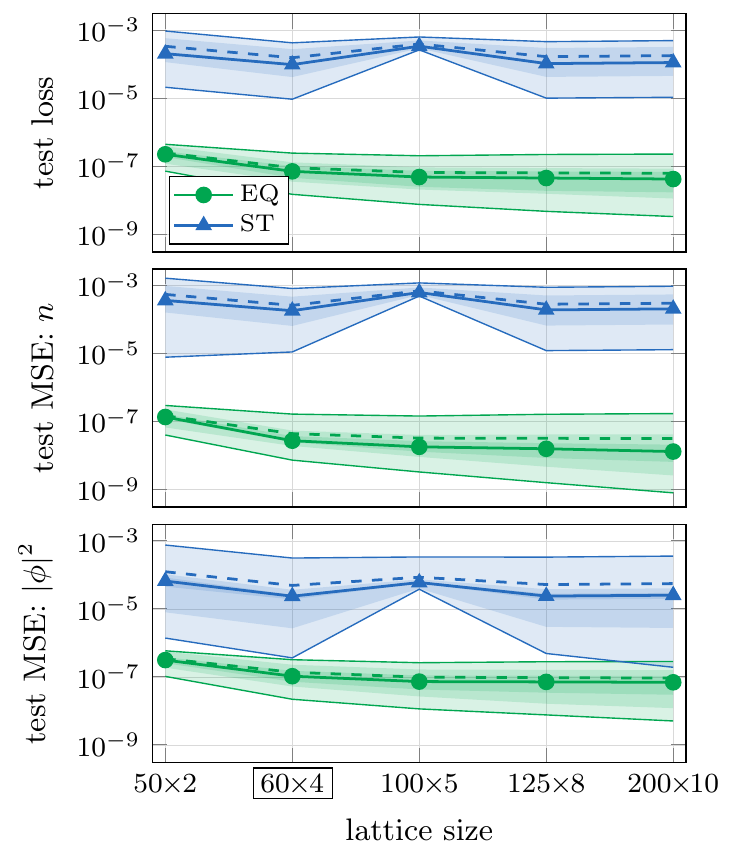}
    \caption{Comparison of the test loss as a function of the lattice size. The best EQ and ST architectures are tested on several lattice sizes, while having been trained on only one specific value ($60\times4$). Image from~\cite{Bulusu:2021rqz}.}
    \label{fig:loss_vs_size}
\end{figure}

To further convince the reader, fig.~\ref{fig:loss_vs_size} reports another comparison between architectures. It shows a test on various lattice sizes, which excludes FL architectures from this comparison. The EQ architectures are about three orders of magnitude more accurate than ST at every lattice size, which suggests that the incorporation of translational equivariance is advantageous. To support this even more, in~\cite{Zhou:2019} an FL architecture with about $10^7$ parameters trained on the $200\times10$ lattice and on two values of $\mu$ reached a test loss of $10^{-6}$, which is two orders of magnitude worse than the best model presented here.

\section{Open worm detection}

\begin{figure}[t]
    \centering
    \begin{subfigure}[t]{0.34\textwidth}
    \centering
    \includegraphics[scale=0.7]{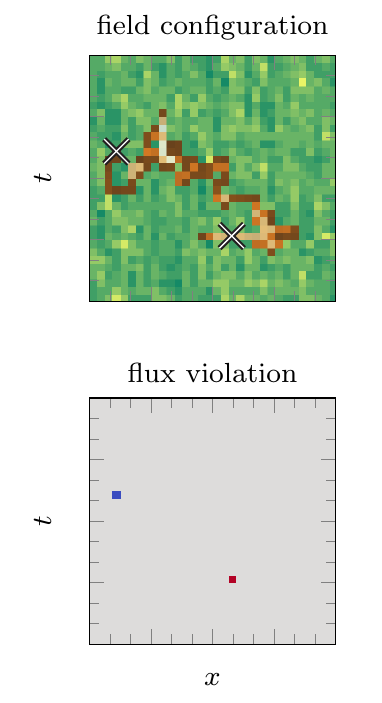}
    \caption{Example field configuration}
    \label{fig:worm}
    \end{subfigure}
    \hfill
    \begin{subfigure}[t]{0.65\textwidth}
    \centering
    \includegraphics[scale=0.7]{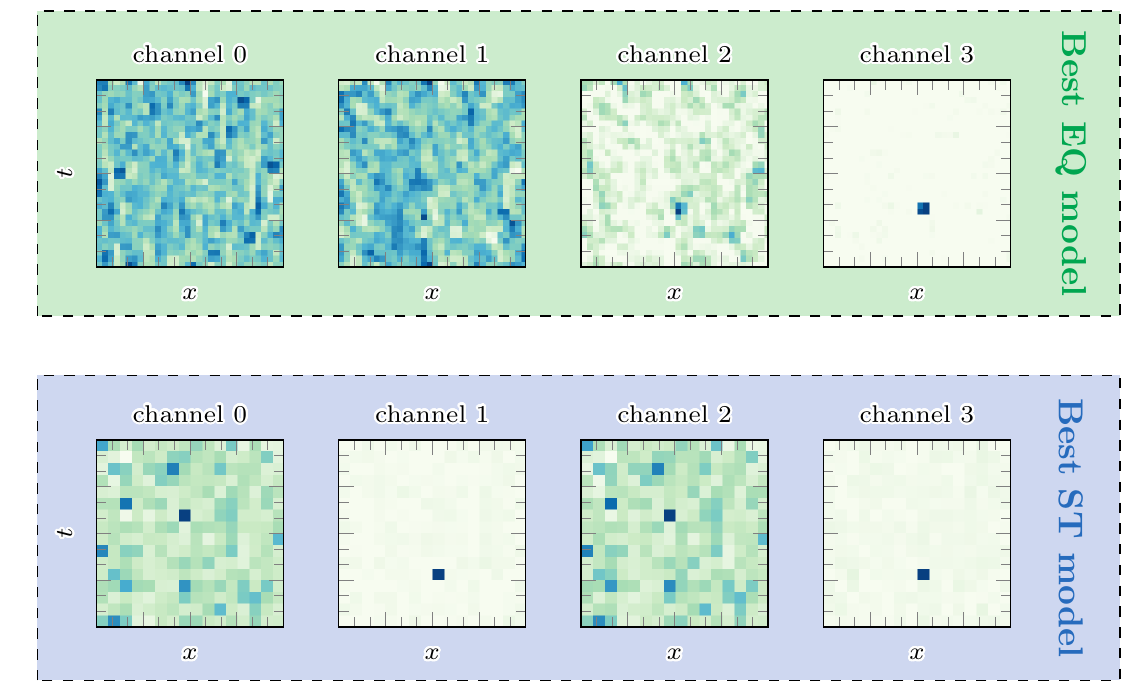}
    \caption{Feature maps of convolutional part in best EQ and ST models}
    \label{fig:flux_detection}
    \end{subfigure}
    \caption{Open worm and flux violation detection. The top plot in (a) is a configuration with an open worm, depicted in brown, on top of a physical configuration. The flux violations at the worm endpoints are visualized in the bottom plot of (a). The open worm detection in the networks is triggered by only one of the two flux violations. Image from~\cite{Bulusu:2021rqz}.}
    \label{fig:flux}
\end{figure}

In the flux representation, the field $k$ satisfies the conservation law $\sum_{\nu} \left( k_{x, \nu} - k_{x - \hat{\nu}, \nu} \right) = 0$. The Prokof'ev-Svistunov algorithm~\cite{Prokofev:2001} is employed to generate configurations that respect such an equation. It proposes local field updates in successive steps on adjacent lattice sites. One can keep track of where the field modifications occur, as is done in fig.~\ref{fig:worm}. The resulting path is referred to as a worm. Its endpoints exhibit flux violations, unless they coincide, in which case the flux conservation is restored at every site. Open worm configurations are characterized by two flux violations, and are therefore unphysical, while in closed worm configurations the flux is conserved everywhere, meaning they are physical. In the previous task, we have used only physical configurations, but here we deliberately create unphysical configurations and demand that the networks distinguish between the two situations.

In order to investigate if the networks are able to generalize, we create a dataset with several combinations of physical parameters and use a small subset in the training process. We choose squared lattices with $N_t=N_x\in\{8,16,32,64\}$; the possible values of the chemical potential are $\mu\in\{1,1.25,1.5\}$; the parameter involving the mass term lies in the set $\eta\in\{4.01,4.04,4.25\}$; the coupling constant is kept fixed at $\lambda=1$. The training subset is given by the two combinations $(\eta,\mu)\in\{(4.01,1.5),(4.25,1)\}$ on $8\times8$ lattices. Open and closed worm configurations are equally represented in the dataset for each parameter combination.

\begin{figure}[t]
\centering
\begin{subfigure}{0.45\textwidth}
\includegraphics[width=5.0cm]{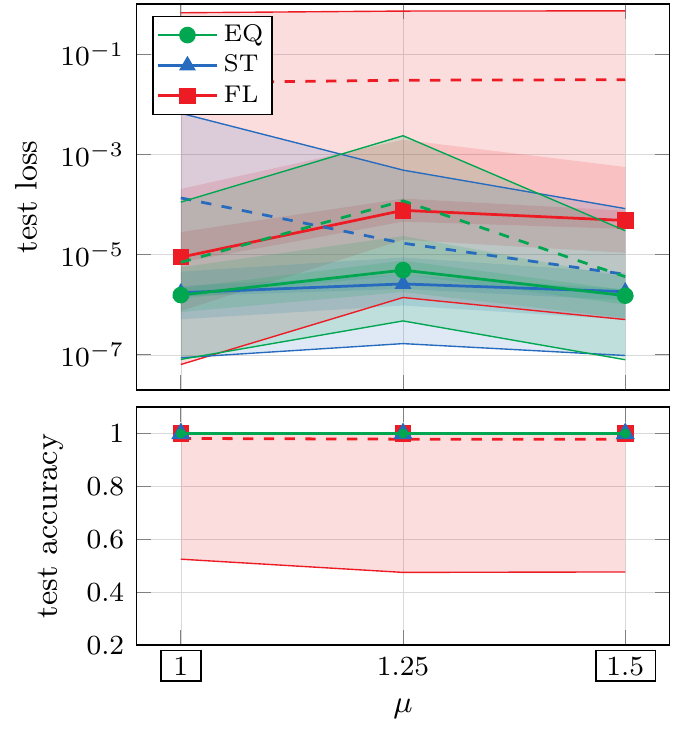}
\caption{Test loss and test accuracy vs.~chemical potential on $8\times8$ lattices for each architecture.}
\label{fig:class_loss_vs_mu}
\end{subfigure}
\hfill
\begin{subfigure}{0.45\textwidth}
\includegraphics[width=5.0cm]{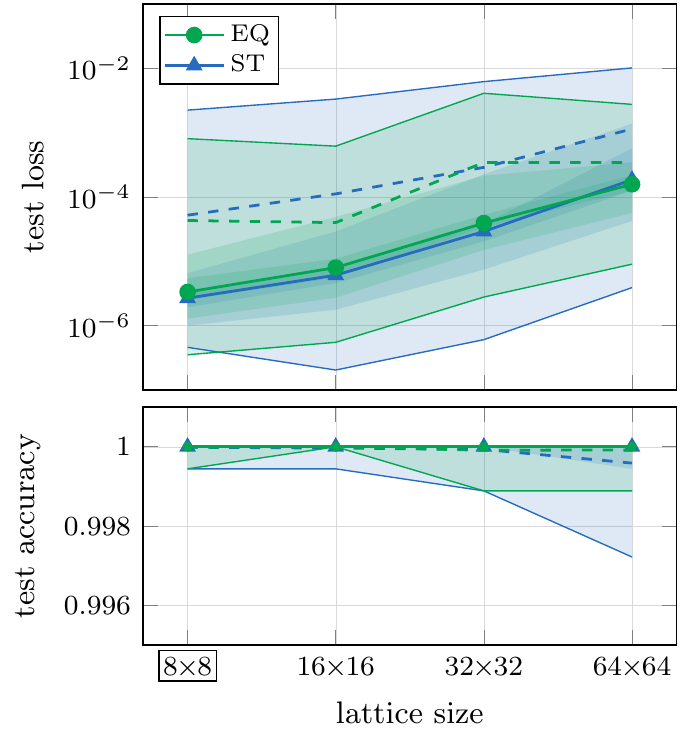}
\caption{Test loss and test accuracy vs.~lattice size for EQ and ST.}
\label{fig:class_loss_vs_size}
\end{subfigure}
\caption{Results for detection of flux violations. The bands delimit the highest and lowest test metrics of 50 models of the architectures suggested by Optuna. In (a), the test loss and the test accuracy are given as a function of the chemical potential on $8\times8$ configurations, whereas in (b) they are reported as functions of the lattice size. Images from~\cite{Bulusu:2021rqz}.}
\label{fig:class_results}
\end{figure}

We adopt the same approach used in the previous task with Optuna, in the attempt to compare the architecture types in the fairest possible way. The hyperparameter search space is slightly modified, for example it is not clear what global pooling type is the most favorable in a classification task, so we include it in the optimization procedure. Monitoring the binary cross entropy loss, Optuna indicates the most promising hyperparameter combination, which is used to train 50 new models using random parameter initialization. The test loss and test accuracy are shown in fig.~\ref{fig:class_results} as functions of the chemical potential and of the lattice size.

While both metrics are not particularly sensitive to changes in the physical parameter $\mu$, larger lattice sizes tend to reduce the model precision. In this case, FL models remain worse than EQ, but ST are more competitive and achieve an accuracy very close to the one of EQ. We also note that, interestingly, successful models learn to detect just one of the two flux violations, as fig.~\ref{fig:flux_detection} suggests.

\section{Multiple open worms counting}

\begin{figure}[t]
\begin{subfigure}{0.47\textwidth}
\includegraphics[width=5.0cm]{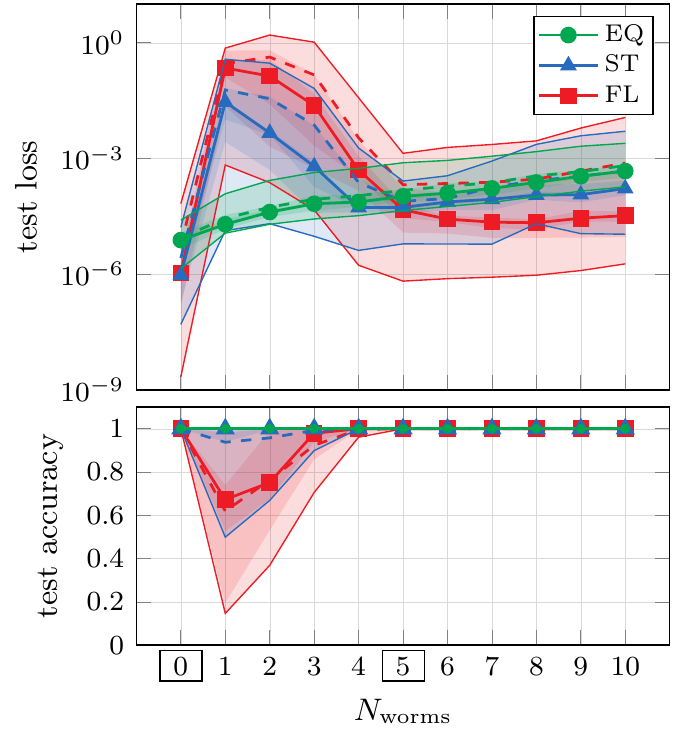}
\caption{Comparison of test loss and test accuracy vs.~number of open worms on $8\times8$ lattices.}
\label{fig:counting_loss_vs_worms}
\end{subfigure}
\hfill
\begin{subfigure}{0.47\textwidth}
\includegraphics[width=5.0cm]{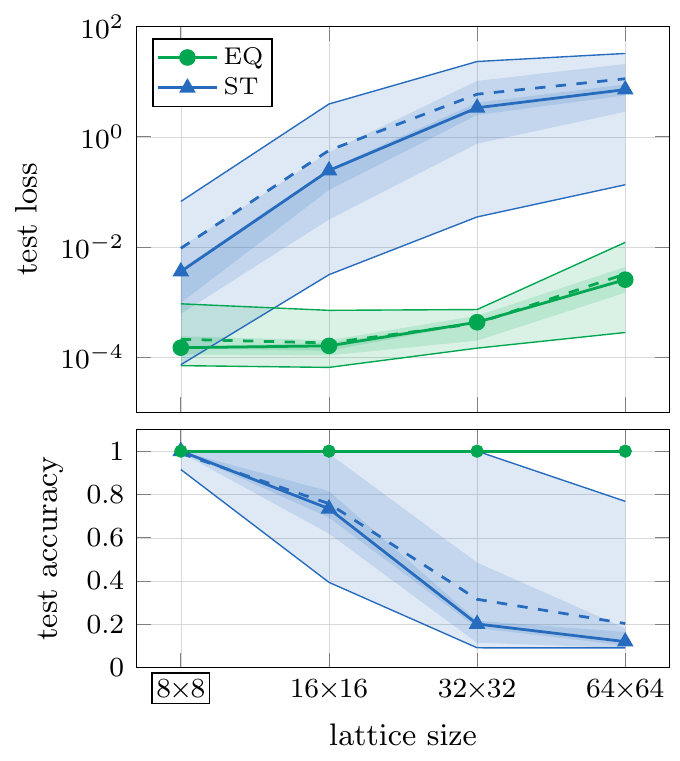}
\caption{Comparison of test loss and test accuracy vs.~lattice size.}
\label{fig:counting_loss_vs_size}
\end{subfigure}
\caption{Results for worm counting. The bands delimit the highest and lowest test metrics of 20 instances of the architectures suggested by Optuna. In (a), the test loss and the test accuracy are given as a function of the chemical potential on $8\times8$ configurations, whereas in (b) they are reported as functions of the lattice size. Images from~\cite{Bulusu:2021rqz}.}
\label{fig:counting_results}
\end{figure}

We revisit the previous task by adding more than one open worm on top of a physical configuration and requiring the networks to assess the right number of open worms. For this, we use the same physical parameters of the last section combined with a number of open worms $N_{\text{worms}}\in\{0,1,\dots,10\}$. The training subset is characterized by the same physical parameters of the single open worm detection, except the number of open worms is 0 and 5.
Using the same strategy with Optuna, we come to the results reported in fig.~\ref{fig:counting_results}. In fig.~\ref{fig:counting_loss_vs_worms} the test loss and test accuracy on $8\times8$ lattices are plotted against the number of open worms. Unlike EQ, which consistently maintains a good performance, ST and FL struggle to correctly predict small numbers of worms. Figure~\ref{fig:counting_loss_vs_size} shows metrics as functions of the lattice size, where EQ clearly outperforms ST, with all of its 20 instances scoring $100\%$ test accuracy.

\section{Conclusions}

An extensive study on the implications of translational equivariance in neural networks was conducted. Three architecture types were designed and tested on three separate tasks, where the reliability of the equivariant type was evident, in particular when it came to the generalization capabilities, both in terms of different physical parameters and of different lattice sizes. The architectures that break translational symmetry led to a much worse performance in almost all cases and are also characterized by inherent drawbacks, such as the impossibility of using an architecture with a flattening layer on different lattice sizes. It is also worth pointing out that Optuna had a preference for small or medium-sized architectures, with a number of parameters between $10^2$ and $10^5$. To conclude, our work indicates that employing translationally equivariant neural networks in translationally symmetric problems can be extremely beneficial.

\vspace{1em}

\begin{acknowledgement}
This work has been supported by the Austrian Science Fund FWF No.~P32446-N27, No.~P28352 and Doctoral program No.~W1252-N27. The Titan\,V GPU used for this research was donated by the NVIDIA Corporation.
\end{acknowledgement}

\bibliography{references}

\end{document}